\documentstyle{mn}

\title[The Energetics and Mass-loss of Mrk~33]
{The Energetics and Mass-loss of the Dwarf Starburst Markarian~33}

\author[L. K. Summers, I. R. Stevens and D. K. Strickland.]{Lesley K.
Summers$^{1}$, Ian R. Stevens$^{1}$ and David K.
Strickland$^{2}$\thanks{{\it Chandra} Fellow.}\\ 
$^{1}$ School of Physics \& Astronomy, University of Birmingham, 
Edgbaston, Birmingham, B15 2TT, UK\\
lks@star.sr.bham.ac.uk; irs@star.sr.bham.ac.uk\\
$^{2}$ Department of Physics \& Astronomy, The Johns Hopkins University,
3400 North Charles Street, Baltimore, MD 21218, U.S.A.\\
dks@pha.jhu.edu}

\date{Accepted .....................; Received .....................; 
in original form .......................}

\begin{document}

\maketitle

\begin{abstract} 

We present {\sl ROSAT} HRI X-ray data and optical imaging of the
important dwarf starburst Markarian~33.  We find an extended, complex,
shell-like morphology in the X-ray emission, with an extent of $\sim
2.3\times 1.9$kpc, coincident with the bright star-forming regions at
the centre of the galaxy.  The physical extent of this X-ray emission
from Mrk~33 is very similar to the observed H$\alpha$ emission, and
suggests that the bulk of the X-ray emission is coming from an expanding
superbubble.

We estimate the age and mass of Mrk~33's starburst to be $5.8$~Myr and
$6.9\times 10^{6}M_{\odot}$ respectively with the energy injection rate
in the central regions of the galaxy being $\sim 10^{41}$~erg~s$^{-1}$,
while the associated mass-loss rate from the star-forming regions is
estimated to be $\sim 0.2M_{\odot}$~yr$^{-1}$. We suggest that the X-ray
emission is predominantly powered by starburst type activity and argue
that a blowout in the form of a galactic wind is the most likely fate
for Mrk~33 resulting in the loss of most of the galaxy's metal-enriched
material and a small fraction ($< 1$ per cent) of the ISM.
\end{abstract}

\begin{keywords}
ISM: jets and outflows -- galaxies: starburst -- galaxies: stellar
content -- stars: Wolf-Rayet -- X-rays: galaxies
\end{keywords}

\section{Introduction}

Blue compact dwarf starbursts belong to the group of galaxies which
exhibit the starburst phenomenon, of intense star formation lasting for
relatively short periods of time ($\leq 10^{8}$ yr), while having
absolute blue magnitudes of $M_{B} \geq -18$.  They are generally low
metallicity objects with $Z$ ranging from $1/30$ to $1/3Z_{\odot}$
(Thuan 1983; Loose \& Thuan 1986), which could be the result of
metal-loss associated with the bursting nature of their star-formation.
Intense periods of star-formation are an important phase of galactic
evolution and can have a profound effect on the galaxies in which they
occur so the study of local dwarf starbursts can give insight into the
fate of similar objects that occurred at high redshifts (e.g. the
numerous faint blue compact objects seen on the Hubble Deep Field
images, Mobashar et al. 1996).  The study of dwarfs is also important as
they are the basic building blocks in the hierarchical merging
cosmological scenario and as such are likely to have harboured the
earliest occurrences of star-formation in the Universe.

The onset of a starburst leads to the formation of OB associations or
even super star-clusters (SSCs) containing many massive stars that have
strong supersonic stellar winds injecting both energy and mass into the
interstellar medium(ISM). After a period of about 3.5 Myr they begin to
give rise to core-collapse supernovae which add to both the energy and
mass loss from the starburst.  The effect of these winds and supernovae
is to sweep-up the ISM and produce shock-heating of both ISM and wind
material leading to the production of an expanding superbubble of hot
gas contained within a shell of cool ISM material.  The formation of
such a structure will be accompanied by extended X-ray emission from the
hot gas ($T\sim 10^{6}-10^{7}$~K) and optical H$\alpha$ line-emission
from the cooler ($T\sim 10^{4}$~K) shell.  The continued expansion of
the superbubble into the halo of the galaxy is due to it having a
greater internal pressure than the surrounding ISM.  The onset of
Rayleigh-Taylor instabilities is likely to lead to the rupture and
subsequent escape of the hot gas in the form of a galactic wind, as seen
most spectacularly in M82 (Strickland, Ponman \& Stevens 1997) and
NGC~253 (Strickland et al., 2000). Typical signatures of such an event
are spurs of X-ray emission coincident with the H$\alpha $ emission
(Heckman et al. 1995) and horse-shoe shaped H$\alpha $ emission with the
open end furthest from the centre of any starburst regions (Marlowe et
al. 1995).  The optical and X-ray phenomena described here have been
observed in several dwarf starburst galaxies. For example, He~2-10
(M\'{e}ndez et al. 1999), NGC~1569 (Heckman et al., 1995) and NGC~5253
(Strickland \& Stevens, 1999). Such scenarios can have a catastrophic
effect on their host galaxies if it results in the loss of a significant
fraction of their mass, leading to expansion of the galaxy, a drop in
its surface brightness and, potentially, a cessation of star-formation.
The rupturing of superbubbles, allowing the venting of hot
metal-enriched gas will effect the galaxy's chemical composition and
evolution.  Such blow-outs could explain the low-metallicities observed
in dwarf galaxies and because of their large numbers, the mass and
energy lost from them during such processes will have had a profound
impact on the state of the intergalactic medium (Dekel \& Silk 1986).

We present here new observations of Mrk~33 (Haro~2, UGC~5720, Arp~233,
IRAS~10293+5439).  This blue compact dwarf galaxy (BCDG) with a mass
$\sim 10^{9}~M_{\odot}$ (Loose \& Thuan 1986) lies at a distance of 22
Mpc (Conti 1991) assuming $H_{0}=75$~km~s$^{-1}$~Mpc$^{-1}$.  At this
distance, $1{''}$ is equivalent to a distance of 0.11~kpc.  Mrk~33 does
not appear to have any obvious companions however there is a very red
galaxian object ($B-R=1.7$ from the USNO A2.0 Catalogue, Monet, 1998)
lying at a projected distance of $\sim 30$~kpc ($4.7{'}$) and at a
position angle of $63^{\circ }$ with respect to Mrk~33's position of
$\alpha=10^{h} 32^{m} 32^{s}$ and $\delta=+54^{\circ } 24{'} 03.5{''}$
(J2000).  Somewhat closer in projection, M$\ddot{o}$llenhoff et
al. (1992), detected a radio source with no obvious optical counterpart
$\sim 3.2$~kpc ($30{''}$) north of Mrk~33. As there is no kinematic data
available for either of these objects, it is impossible to say whether
they have any association with Mrk~33.  Described by Huchra (1977) as
being, `patchy with wisps and a high surface brightness', Mrk~33 was
later given an nE classification by Loose \& Thuan (1985; 1986).  Due to
its apparently straight-forward morphology, Mrk~33 is the prototype of
this class of object and as such gives an interesting and important
snapshot of this intriguing phase of galaxy evolution. The galaxy's blue
colour is emphasised by its B magnitude ($m_{B} \sim 13.40$, de
Vaucouleurs et al., 1991) and its $(B - V)$ colour of 0.36 (Thuan \&
Martin 1981), which is bluer than other starburst galaxies
(e.g. NGC~5253 $(B - V) \sim 0.44$; NGC~1569 $(B - V) \sim 0.77$; M82
$(B - V) \sim 0.87$ all values quoted are from the Sky Catalogue 2000.0
Vol.  2, (Hirshfeld \& Sinnott 1997)).  Compared with other BCDGs, which
have metallicities $\sim 0.02-0.20$~solar, Mrk~33, with a value of $\sim
0.3$~solar (Legrand et al. 1997), seems to have a somewhat high value.

Radio observations of Mrk33 show it to have a steep spectrum with a
spectral index of $\alpha = -0.82\pm 0.12$ (Klein et al. 1984),
suggesting the presence of non-thermal sources such as supernova
remnants. More recent results from Beck et al. (2000) show the presence
of both thermal and non-thermal emission with the former coming mainly
from the outer regions of the galaxy and not from point sources, while
the latter comes mainly from the inner regions.  At shorter wavelengths,
the far-infrared luminosity of Mrk~33 from $40-120\mu$m is $L_{FIR}\sim
1.4\times 10^{43}$~erg~s$^{-1}$ (calculated as the average value quoted
from Thuan \& Sauvage 1992, Melisse \& Israel 1994 and Stevens \&
Strickland 1998).

Optical and UV observations of Mrk~33 along with spectral synthesis
techniques show that the galaxy has experienced at least two other
episodes of intense star-formation prior to the current one, (Loose \&
Thuan 1986; Fanelli et al. 1988).  The most prominent emission feature
in the optical spectrum is that of the H$\alpha$+N[II] blend and Legrand
et al.'s (1997) study of this emission line found evidence of a
partially ionised wind out-flowing from the star-forming region at $\sim
200$~km~s$^{-1}$.  The Ly$\alpha$ spectrum has a P-Cygni type profile
which also contains blue-shifted Ly$\alpha$ emission indicative of the
outflow of partially ionised gas at a speed $\sim 200$~km~s$^{-1}$
(Lequeux et al. 1995).  Tenorio-Tagle et al. (1999) have predicted such
profiles as the result of an expanding ionization front trapped within
the recombining shell of an expanding superbubble, when the bubble has
an age $\sim 5$~Myr.

A value for the soft X-ray luminosity of Mrk~33, in the $0.1-2.4$~keV
band, was obtained from {\sl ROSAT} PSPC data by Stevens \& Strickland
(1998) in their survey of X-ray emission from Wolf-Rayet galaxies.  The
value quoted was $L_{X}=1.4\times 10^{40}$~erg~s$^{-1}$ resulting from a
best fit to the spectrum of a single-temperature Raymond-Smith model
with $kT=0.36$~keV and $N_{H}=2.2\times 10^{21}$ cm$^{-2}$. The quality
of the X-ray spectral data was poor, and to obtain this fit, the
metallicity was fixed at 0.1 solar, somewhat lower than the 0.3 solar
quoted earlier. Consequently, there are substantial uncertainties in the
spectral fit and in particular on the value of $N_{H}$. Hence, when the
extreme values of $N_{H}$ were used, the intrinsic X-ray luminosity was
only constrained to be in the range $2\times 10^{39}-1.4\times
10^{42}$~erg~s$^{-1}$, although the uncertainty on the observed value
will be much less than this range.

In Section~2, we describe the new observations, firstly the X-ray
observations with the {\sl ROSAT} HRI instrument and secondly $B$ and
$R$ band optical imaging with the 1.0~m Jacobus Kapteyn Telescope (JKT)
on La Palma.  The results obtained from these observations are detailed
in Section~3 while the implications of the optical results and the
energetics of the starburst are discussed in Section~4 with a view to
determining the mass-loss rate from the starburst region, the energy
injection rate into the ISM and their possible implications on the
subsequent evolution of Mrk~33.  Our main conclusions are summarised in
Section~5.

\section{Observations and Analysis}
\begin{figure*}
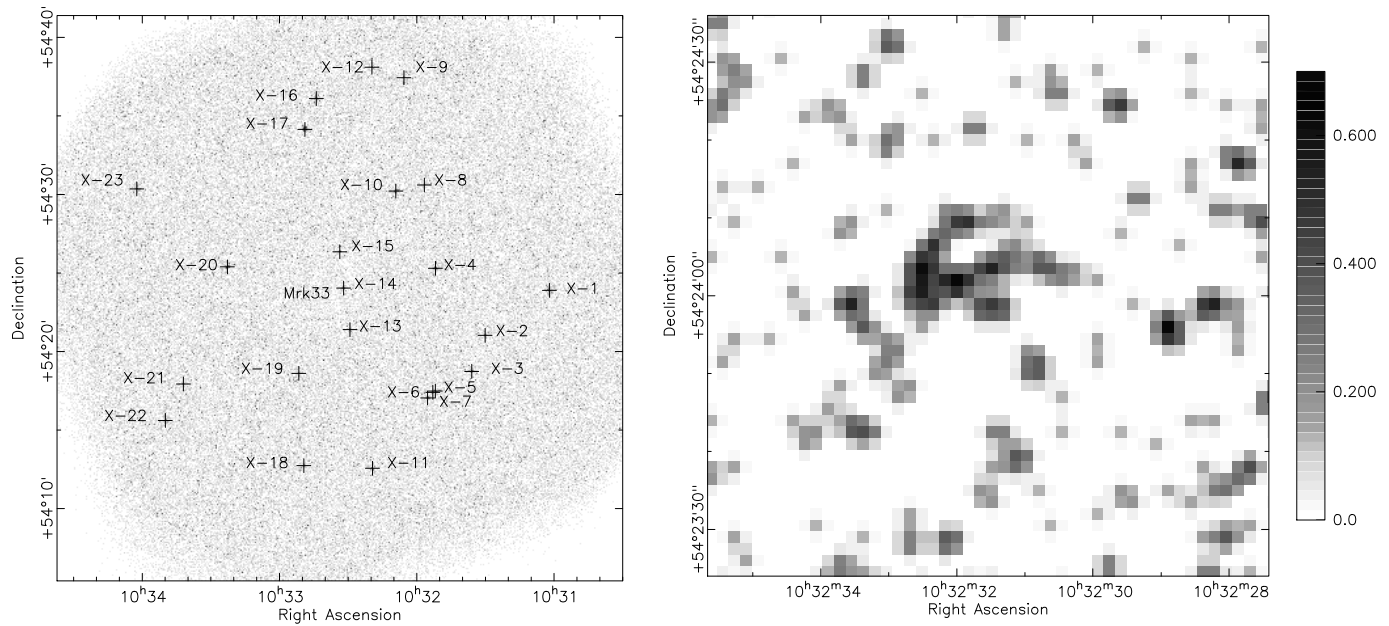

\vspace{10cm}
\includegraphics{968fig1a.ps}
\includegraphics{968fig1b.ps}
\caption{The {\sl ROSAT} HRI data on the Mrk~33 field.  Left: A low
resolution image of the 47.7ks {\sl ROSAT} HRI observation. The image
has been marked with the 23 sources detected with a significance $\geq 4
\sigma$ above background. The full listing of these sources is given in
Appendix A and their positions are only accurate to within $10{''}$.
Right: A high-resolution, background subtracted, HRI X-ray image of
Mrk~33.  The key shows the counts pixel$^{-1}$ above the background
level.  North is to the top and East to the left.}
\label{ImXC}
\end{figure*} 

\subsection{New HRI Observation}

Mrk~33 was observed with the {\sl ROSAT} HRI on April $29^{th}$ 1997 for
47676s.  By using two microchannel plates and a crossed grid position
read out system, the HRI produces images at its prime focus with a
spatial resolution $\sim 5{''}$. The low number (16) of pulse height
analyzer (PHA) channels mean that its spectral analysis capabilities are
limited (David et al., 1995), however it is possible to calculate
hardness ratios for objects (See later). The raw data set was obtained
from the Leicester Data Archive (LEDAS) and the data analysis was
performed using the Starlink ASTERIX X-ray analysis package.  After
screening for bad data, no correction to the data needed to be made. Two
initial images were produced from the data, the first, for point source
searching of the entire field, was $0.6^{\circ }\times 0.6^{\circ }$
square with the data binned in 5 arcsec square pixels, while the second,
for detailed study of the region around Mrk~33, was of higher
resolution, being $0.2^{\circ}\times 0.2^{\circ }$ square with the data
binned in 1.5 arcsec square pixels. In both cases, only the photons
received in PHA channels 3 - 8 were used so as to reduce the
contribution from background sources (e.g. the detector's internal
background, the X-ray background and charged particles). A constant
background model was constructed for each of these images using a source
free annulus with $r=0.05-0.08^{\circ }$ centred on the centre of the
HRI field of view, which lies $\sim 40{''}$ SSE of Mrk~33.  The count
rate obtained for each of these background models was $(2.93\pm
0.25)\times 10^{-3}$~ct~s$^{-1}$~arcmin$^{-2}$.  This value was then
subtracted from each image and the resulting background subtracted
images were then smoothed using a Gaussian with $\sigma =5{''}$, in the
low resolution case and $3{''}$ in the high resolution case.  The images
from both of these procedures are shown in Fig.~\ref{ImXC}.

\subsubsection{Point Source Searching}

The ASTERIX maximum likelihood point source searching algorithm, PSS
(Allan 1993), was used to detect point sources with a significance
greater than $4\sigma$ above the background.  Although 22 sources,
excluding Mrk~33, were detected in the HRI field of view, none of them,
within a radius of $20{''}$, could be identified with known point
sources in the SIMBAD database or in the USNO-A2.0 catalogue down to an
apparent magnitude of 17.5 (Monet 1998).  As a result, no correction
could be made for any pointing errors and so the positions of these
objects are accurate only to within $10{''}$ (David et al. 1995).  The
23 sources, including Mrk~33, are shown in Fig.~\ref{ImXC} superimposed
on the low-resolution X-ray image and for completeness, the full listing
of all 23 point sources in the field of view is presented in Appendix~A.

\subsection{New JKT imaging}

\begin{figure*}
\vspace{10cm}
\includegraphics{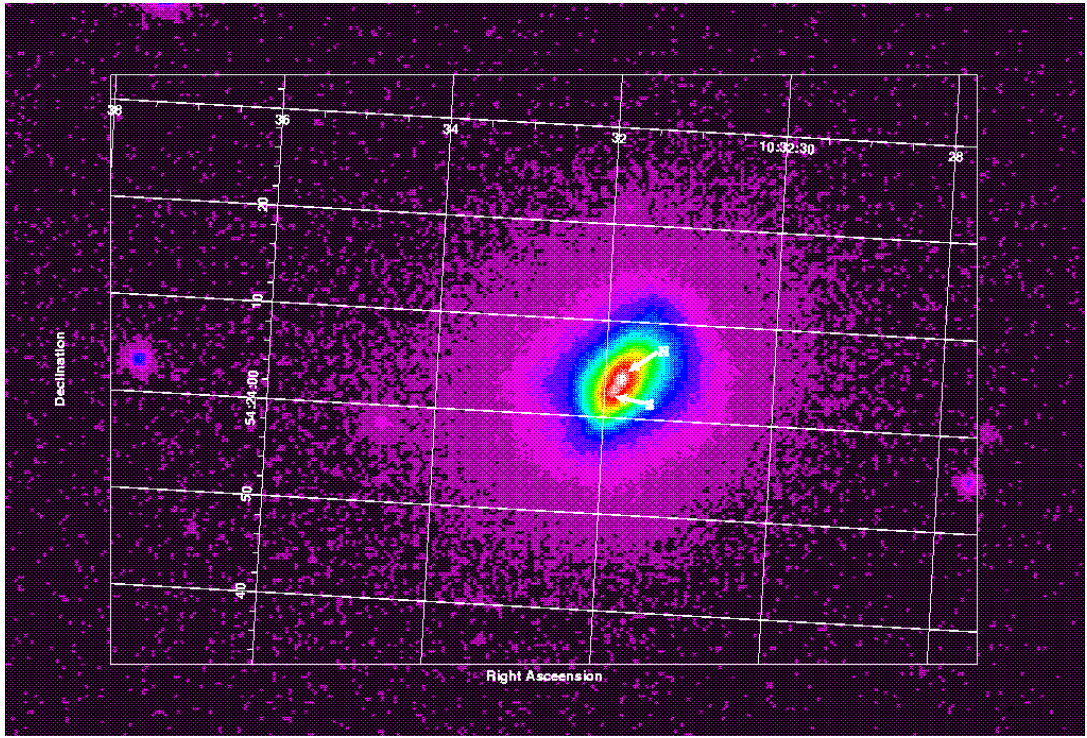} 
\caption{The JKT $B$ band image of Mrk~33 with the two peaks of emission 
from the star-forming regions labelled N and S.}  
\label{Blue}
\end{figure*}

Mrk~33 was observed using the 1.0~m JKT, on La Palma, on the $2^{nd}$
January 2000.  Seven $B$ band and seven $R$ band images, each with an
exposure time of 300s were produced using Harris $B$ and $R$ filters and
the {\sl SITe2} CCD.  One of the $B$ band images was subsequently
rejected because of contamination in the same area of the image as
Mrk~33.  In addition to these target images, bias frames, flat-fields,
and calibration images were also obtained.  The images were processed
using the IRAF CCD processing software, {\sl ccdproc} in combination
with its image processing software, {\sl images}.  After bias
subtraction and flat-fielding, the target frames were weighted and
aligned before being combined.  The weighting was achieved using the
IRAF digiphot aperture photometry software, {\sl apphot} and the
photometric calibration software {\sl photcal}, on the calibration
images to find the atmospheric extinction and colour corrections to be
applied to each target image.  The required standard star magnitudes and
colours were taken from the Landolt catalogue (Landolt 1992).  The
resulting $B$ and $R$ band images are shown in Fig.~\ref{Blue} and
Fig.~\ref{ContB} respectively.

\section{Results}

\subsection{JKT Data}

\begin{figure}
\vspace{7.0cm}
\includegraphics{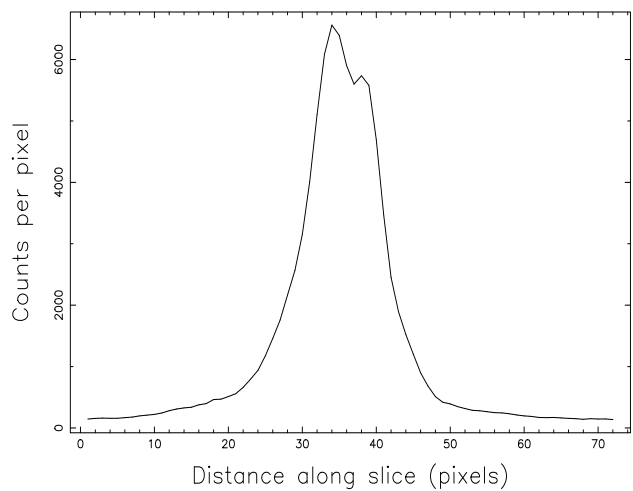} 
\caption{Slice through the blue image of Mrk33 showing the number of counts
per pixel in a slice running from NW to SE with 10~pixels $= 5^{''}$. The 
positions of the peaks
and the ends of the cut are given in the text.}
\label{gslice}
\end{figure}

The $B$ band image of Mrk~33 shown in Fig.~\ref{Blue} confirms that the
intense star-formation occurring in the galaxy is confined to the nuclear
region and suggests that the radius of this region's extent is of $\sim
10-15{''}$ from the peak of the optical emission (corresponding to $\sim
1-1.5$~kpc for an assumed distance of 22~Mpc).  In comparison, the low
surface brightness extent of the galaxy can be seen to extend out to $\sim
1{'}$ ($\sim 6$~kpc) in diameter. Fig.~\ref{gslice} shows the counts per
pixel for a slice across the $B$ band image running from the NW to the SE
of the galaxy from a position of $\alpha=10^{h} 32^{m} 31^{s}$ and
$\delta=+54^{\circ} 24{'} 16{''}$ (J2000) to a position of $\alpha=10^{h}
32^{m} 32.7^{s}$ and $\delta=+54^{\circ} 23{'} 51{''}$ (J2000). The image
shows 2 peaks of emission, labelled N and S in Fig.~\ref{Blue}, occurring
at positions of $\alpha=10^{h} 32^{m} 31.8^{s}$ and $\delta=+54^{\circ}
24{'} 04{''}$ (J2000) and $\alpha=10^{h} 32^{m} 31.9^{s}$ and
$\delta=+54^{\circ} 24{'} 03{''}$ (J2000) respectively, which are extended
over a distance of at least $10{''}$.

\subsubsection{Surface Brightness Radial Profiles}

From the optical images, surface brightness profiles have been obtained
using the isophotal plotting capabilities of IRAF.  The
foreground/background sources that are evident on the $B$ and $R$ band
images were subtracted from the images before the fitting of elliptical
isophotes was carried out.  The ellipses were allowed to vary in both
ellipticity (with the ellipticity defined as $e=1-b/a$, with $a$ and $b$
the fitted semi-major and semi-minor axes respectively) and position
angle, while the ellipse centre was fixed on the optical centre of the
most northern of the optical peaks seen on the JKT images.

\begin{figure*}
\vspace{6.0cm}
\includegraphics{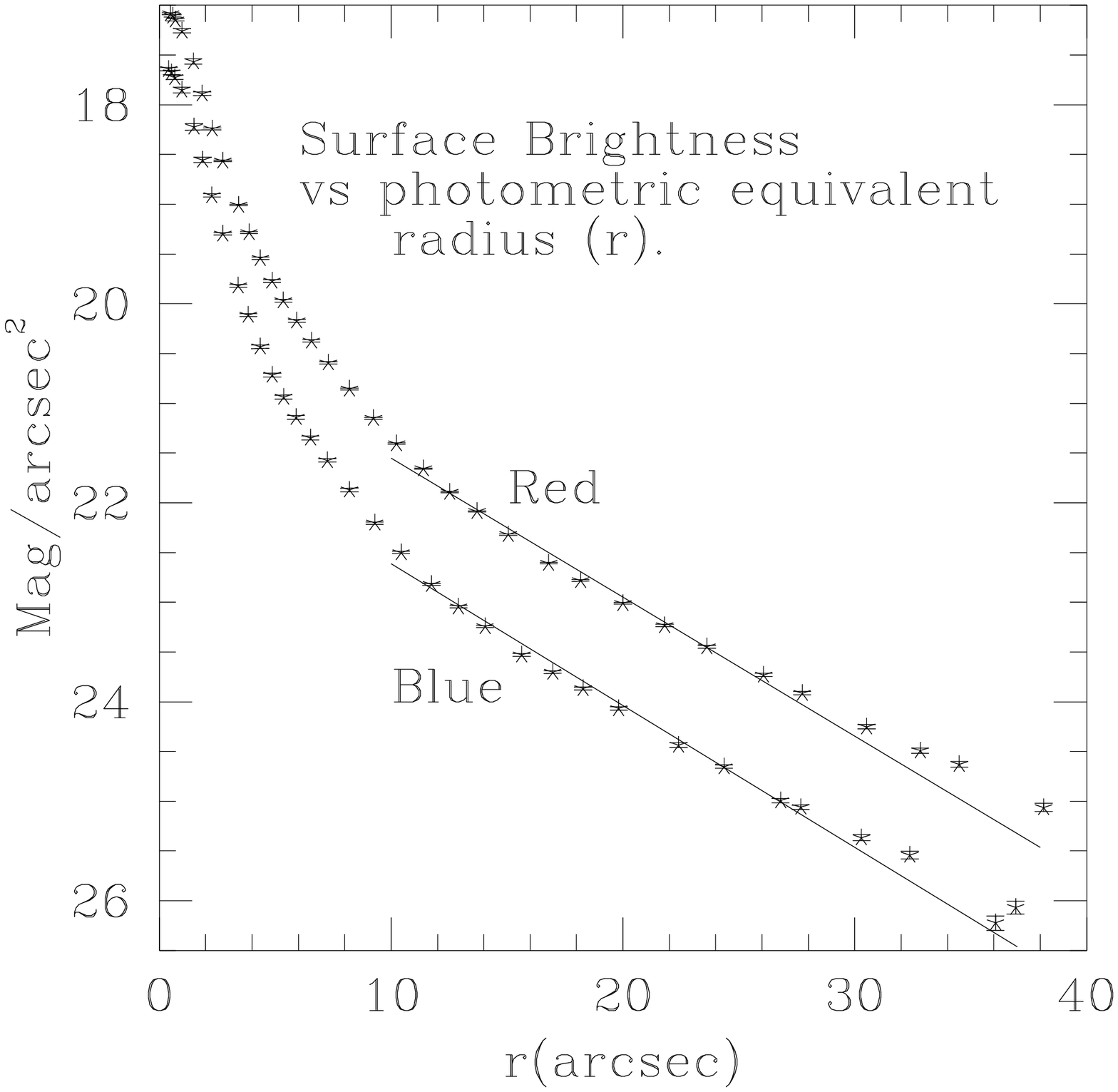} 
\includegraphics{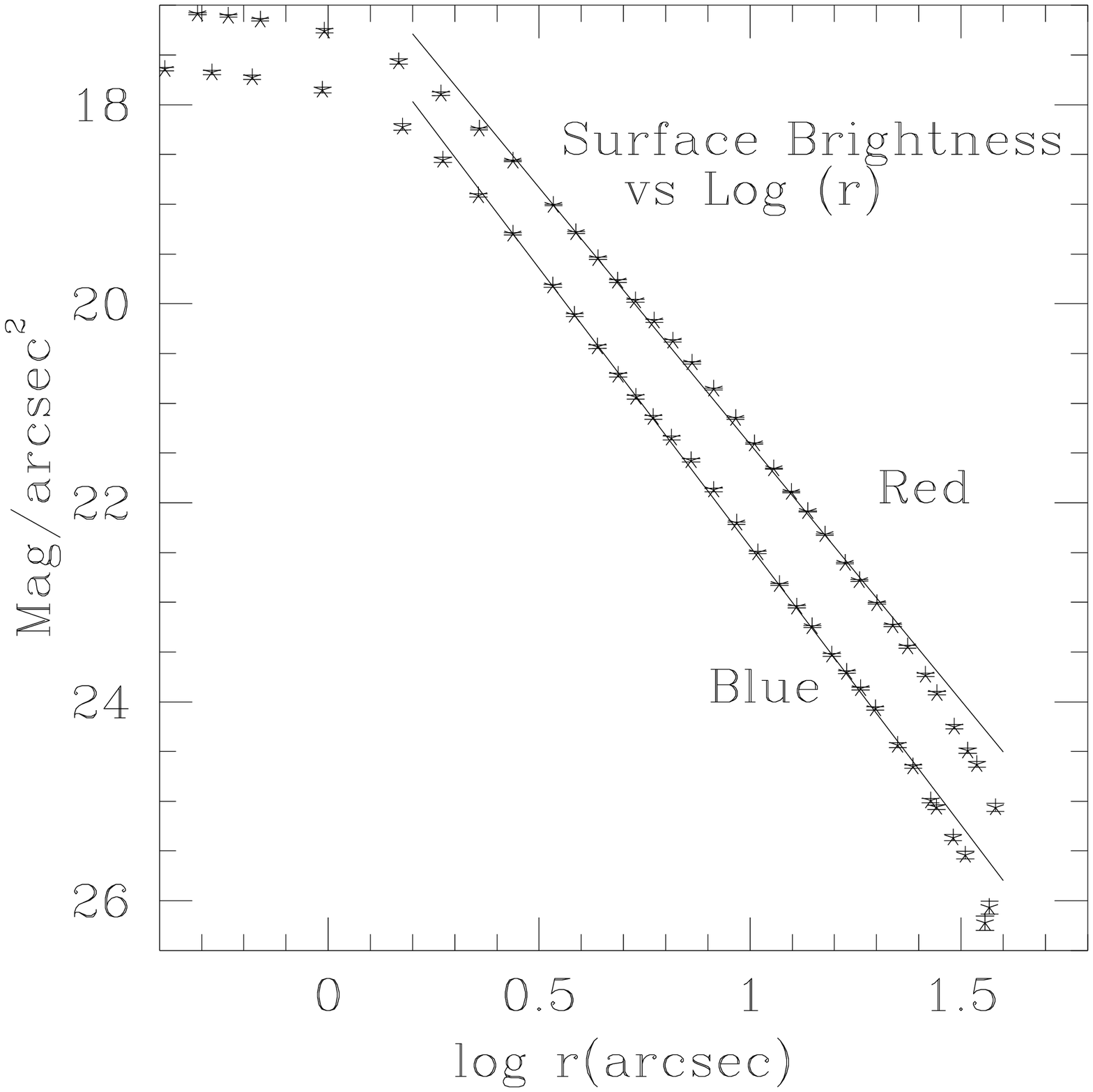} 
\includegraphics{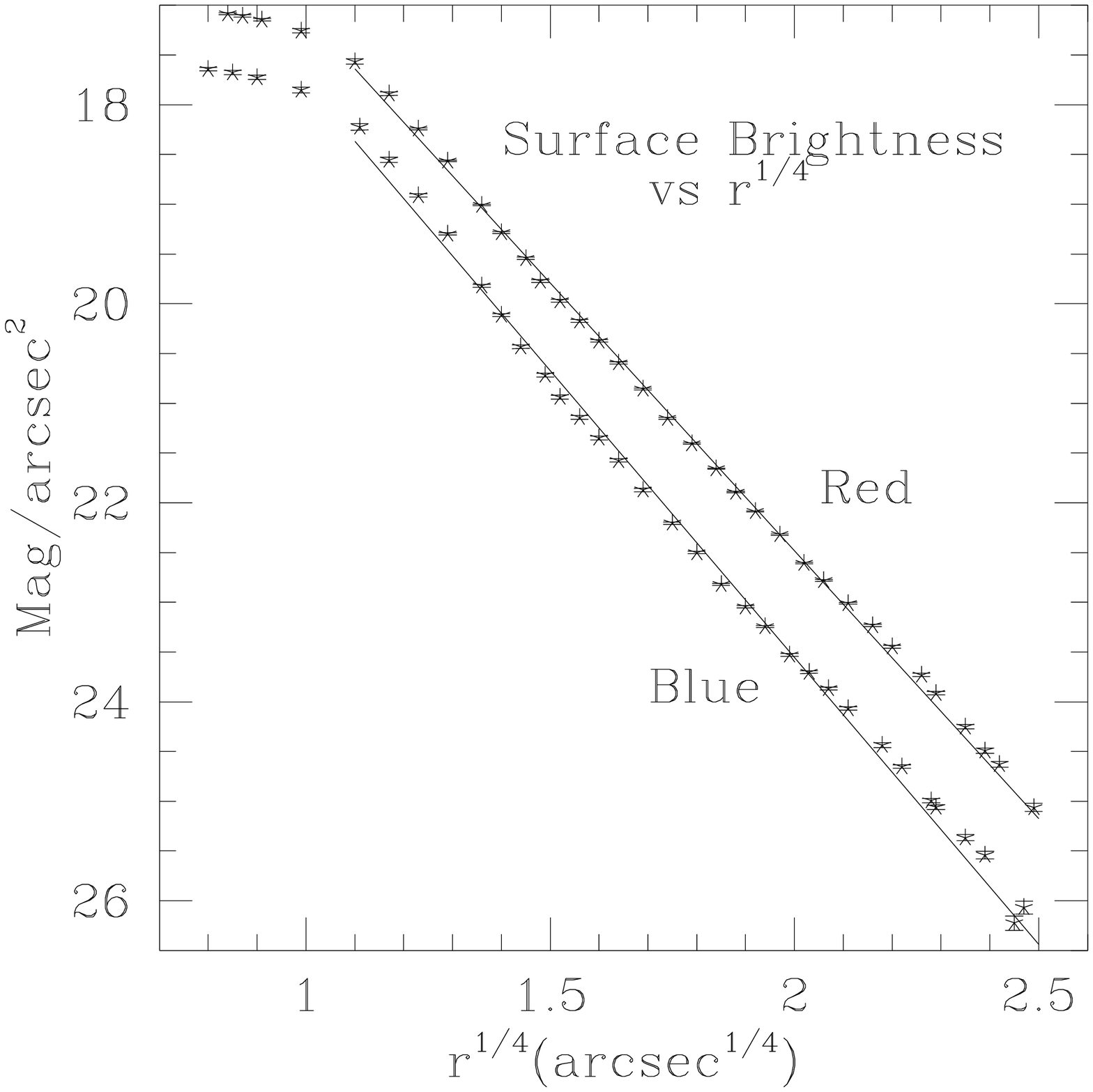} 
\caption{Left: Radial surface brightness profile of Mrk~33 for the JKT B
and $R$ band data.  Centre: Radial surface brightness profile plotted 
against $\log r$. Right: Radial surface brightness profile plotted 
against $r^{1/4}$. Details of the best fit lines are given in the text.}
\label{rsbp}
\end{figure*}

The $B$ and $R$ band surface brightness profiles are plotted in
Fig.~\ref{rsbp}. Following Loose \& Thuan (1996) these have been plotted
in three ways.  The first plot shows the surface brightness against
photometric equivalent radius ($r=\sqrt{ab}$), a plot that is indicative
of an exponential law ($I(r)=I_{0}e^{-ar}$) when giving a straight line
and one which is characteristic of a disk system.  The best fit lines
shown on the plot are for the data with $r>10{''}$ and have slopes of
$-0.057\pm 0.001$ and $-0.056\pm 0.001$ for the blue and red data
respectively.  The second plot shows surface brightness against log
($r$) where a straight line is indicative of a power law
($I(r)=I_{0}r^{-n}$) and the third shows surface brightness against
$r^{1/4}$.  In the last case, a straight line is indicative of a de
Vacouleurs law of the form $I(r)=I_{e}e^{-7.67[(r/r_{e})^{1/4} - 1]}$
which is characteristic of spheroidal systems where, $r_{e}$ is the
effective radius containing half the light emitted from the system and
$I_{e}$ is the effective surface brightness at that radius.  For the
second and third plots in Fig.~\ref{rsbp} the best fit lines shown are
only for the data with $r\geq 1.5{''}$. In the log ($r$) plot, the
slopes of the lines are $-2.24\pm 0.01$ and $-2.06\pm 0.01$ for the blue
and red data respectively. In the case of the $r^{1/4}$ plot they are
$-2.31\pm 0.01$ and $-2.15\pm 0.01$ again for the blue and red data
respectively.

\begin{figure*}
\vspace{6.5cm}
\includegraphics{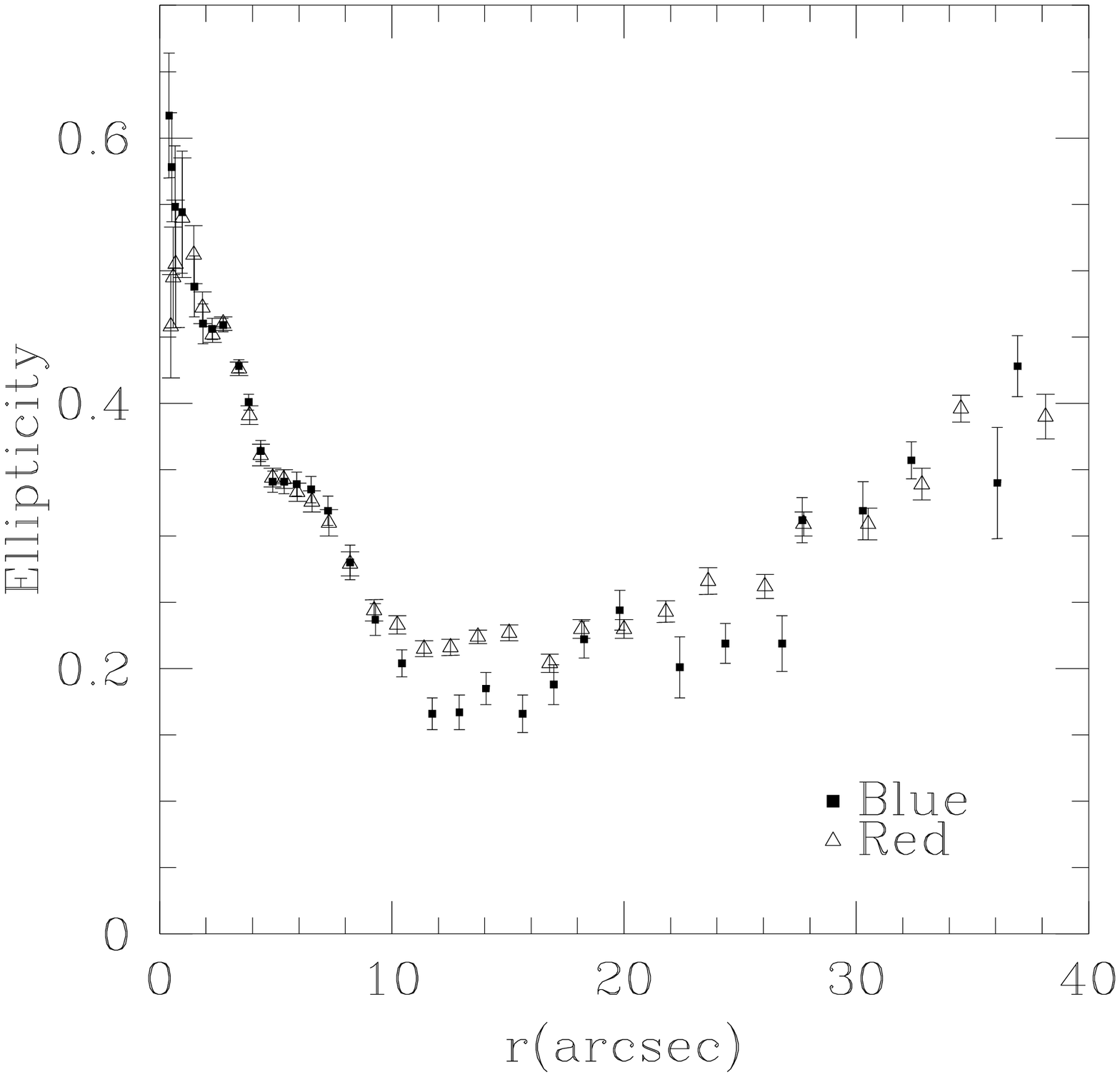} 
\includegraphics{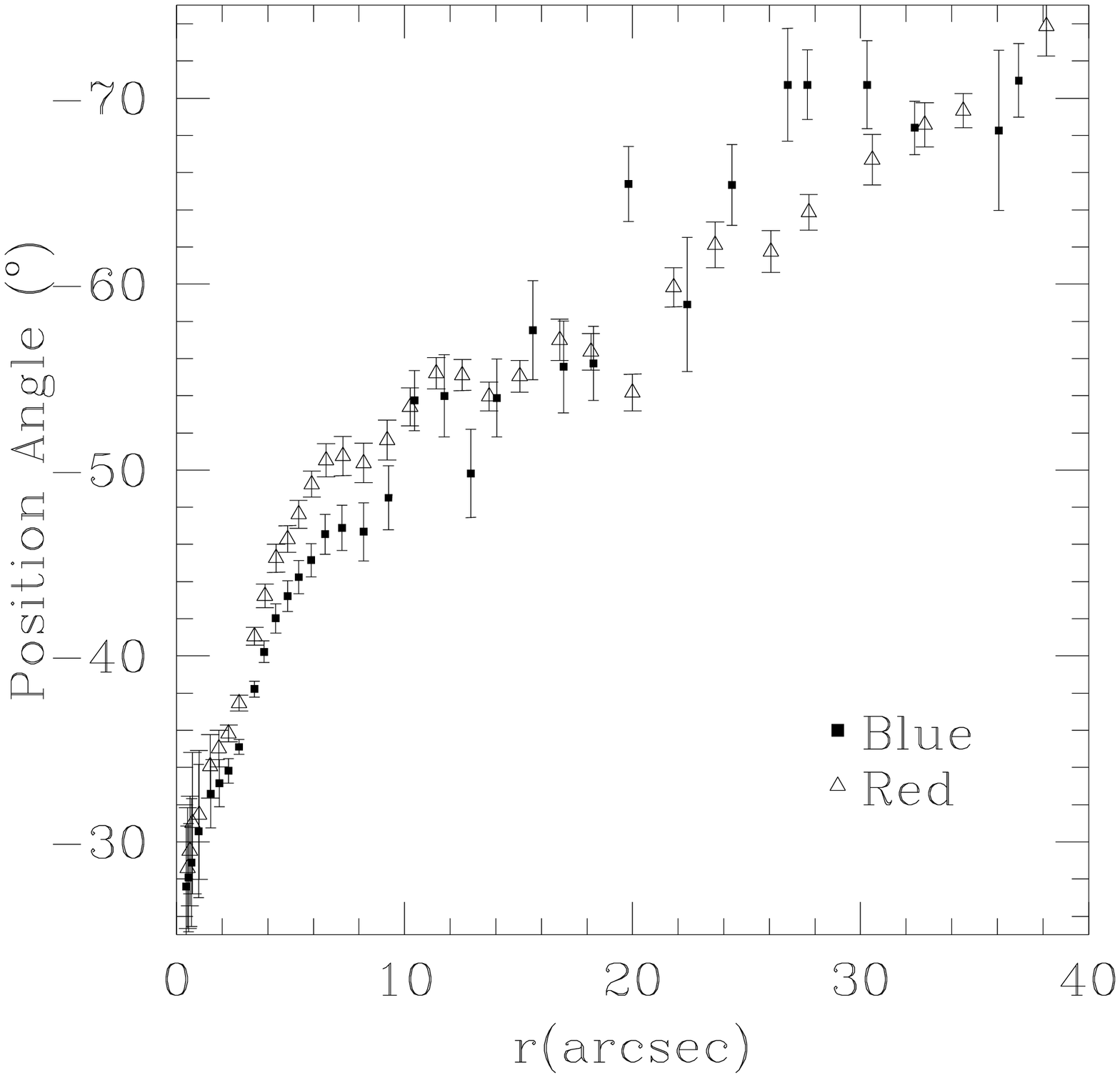} 
\includegraphics{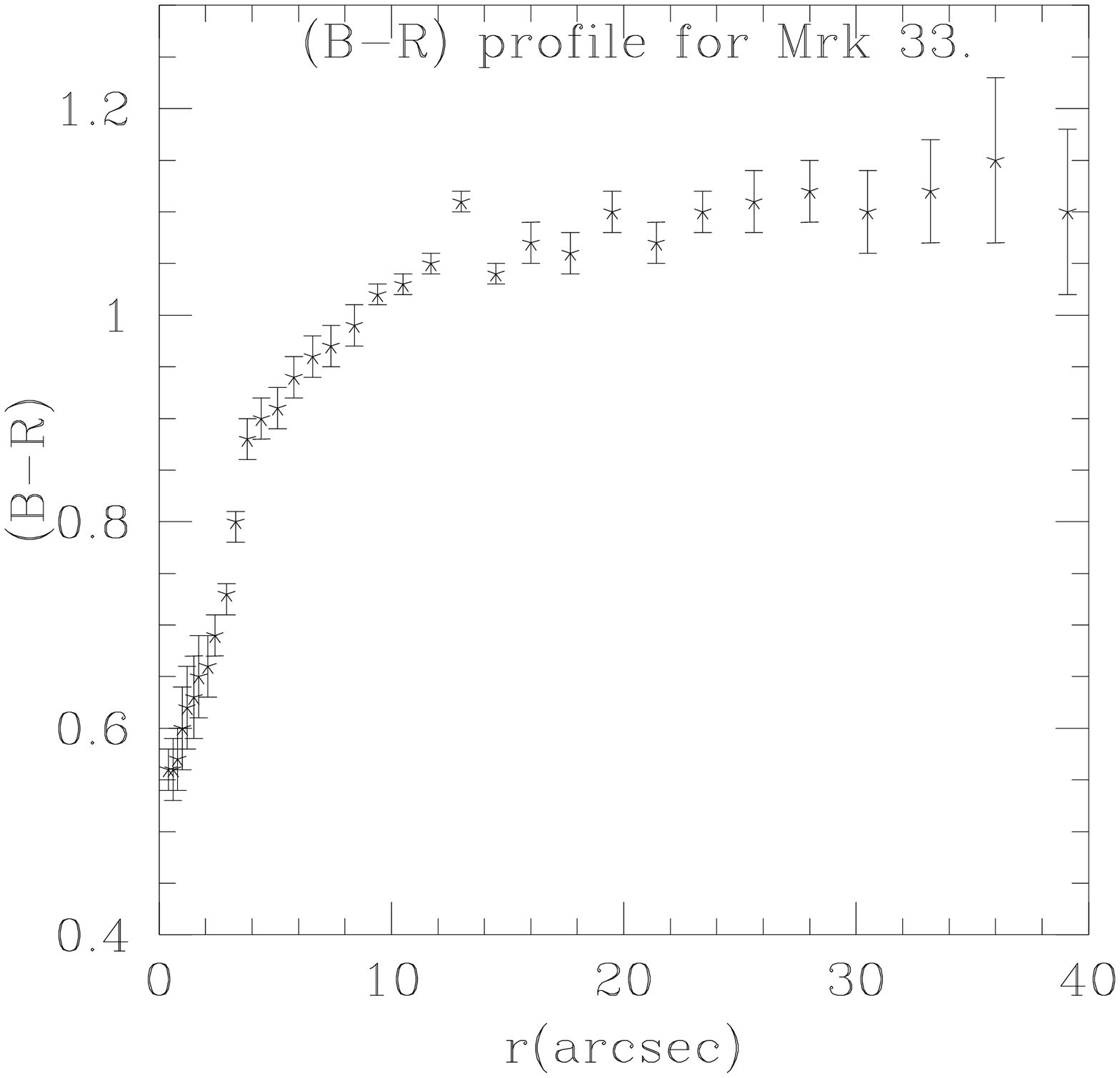} 
\caption{Left: Variation of the ellipticity of Mrk~33, defined as
$e=1-b/a$, for isophotes out to a radius of $40{''}$ ($\sim 4.3$~kpc) for
both the $B$ and $R$ band observations.  Centre: Variation of the position
angle of Mrk~33's isophotes also out to a distance of $40{''}$.  Right:
Variation of the $B-R$ colour of Mrk~33 out to a distance of $40{''}$.}
\label{ell}
\end{figure*}

The ellipticity and position angle profiles are shown in
Fig.~\ref{ell}. The ellipticity does not show a smooth variation from
the centre out but this is probably due to the starburst region
affecting the shape of the isophotes in the central region of the
galaxy. The position angle steadily increases with radius suggesting an
isophotal twist. In addition to these basic profiles, photometry was
performed on both images, resulting in apparent magnitudes of
$m_{B}=13.5$ and $m_{R}=12.6$ for an aperture corresponding to a radius
of $33{''}$ and the $B-R$ colour profile was calculated (also shown in
Fig.~\ref{ell}).  As expected it shows that the galaxy is bluer in the
centre than in the outer regions, reflecting the presence of lots of
young blue stars in the central starburst region.  The uncertainties
shown result from the root mean square fluctuation in the intensity of
the data in the pixels along the path of each individual elliptical
isophot. The implications of these results will be discussed further in
Section~4.

\subsection{HRI Data}

\subsubsection{X-ray Contour Plots and Optical Images}

\begin{figure*}
\vspace{10cm}
\includegraphics{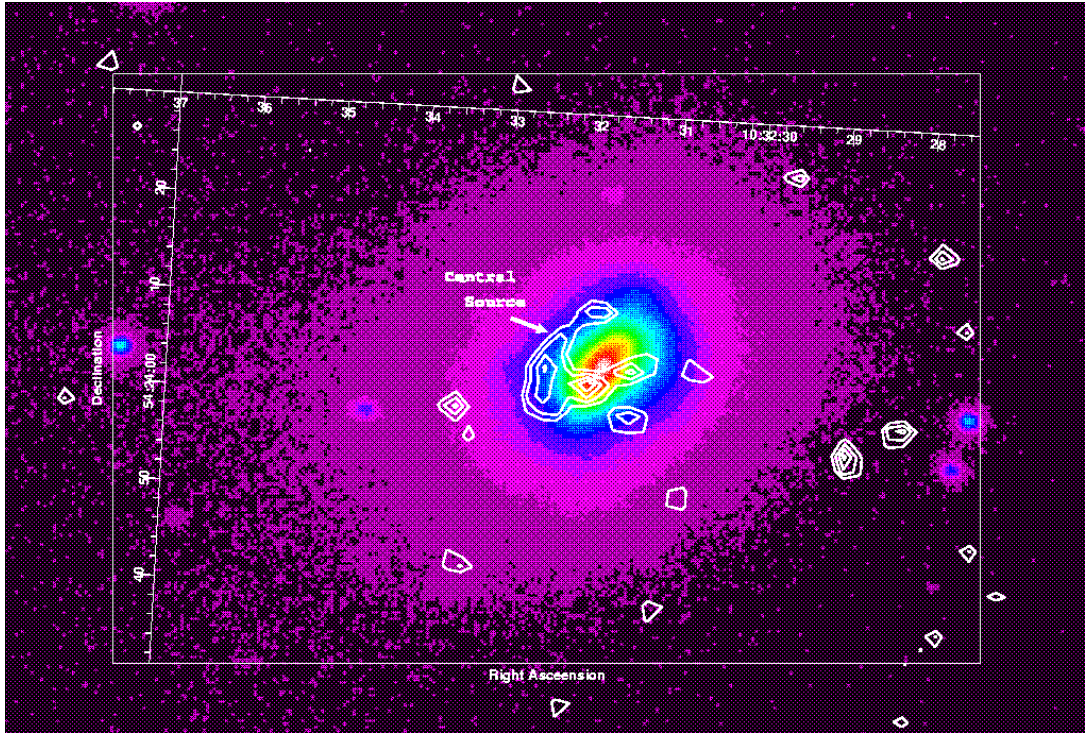}
\caption{High-resolution X-ray contour image overlaid on the $R$ band JKT 
image of Mrk~33. The pixel size is $1.5{''}$ smoothed with a Gaussian 
having a $\sigma$ of $3{''}$ and the contour levels start at $1.01\times 
10^{-2}$~ct~s$^{-1}$~arcmin$^{-2}$ and increase linearly in steps of 
$0.33\times 10^{-2}$~ct~s$^{-1}$~arcmin$^{-2}$}
\label{ContB}
\end{figure*}

X-ray contours from the high resolution X-ray image smoothed using a
Gaussian with $\sigma =3{''}$ have been overlaid on the $R$ band image
of Mrk~33. The resulting image is shown in Fig.~\ref{ContB}. The image
shows a central extended source of X-ray emission and what could be
several point sources of emission lying within a projected distance of
$33{''}$ from its centre at $\alpha=10^{h} 32^{m} 32.2^{s}$ and
$\delta=+54^{\circ} 24{'} 03{''}$. This projected distance was chosen
because it is equivalent to the radius of the circular aperture which
corresponded to half the average $D_{25}$ value, calculated from Thuan
\& Martin 1981, Gordon \& Gottesman 1981, Loose \& Thuan 1986, Melisse
\& Israel 1994 and Thuan \& Sauvage 1992) and used in the optical
photometry.

The extended source is of the order of $21{''}$ ($\sim 2.3$~kpc) in
extent when measured from NNW to SSE and of the order of $17{''}$ ($\sim
1.9$~kpc) in extent from E to W.  Fig.~\ref{ContB} shows that as well as
being extended, this source appears to have a blobby nature to its
emission.

\subsubsection{Emission, Flux and Luminosity from Mrk~33's Extended X-ray 
Source.}

The high-resolution image shown in Fig.~\ref{ImXC} was used to compare
the X-ray photons in channels $3-5$ (softest photons) with those in
channels $6-8$ (hardest photons). No photons were detected from the
extended source in channels $6-8$ hence it is very soft and the hardness
ratio for the region of radius $33{''}$ centred on the extended source,
defined as the number of counts in channels $(6 - 8)/(3 - 5)$, is $0.11
\pm 0.05$, which is softer than most of the other 22 X-ray sources
detected in the HRI field of view (See Appendix A). The count rate for
the extended source is $(4.5\pm 1.0) \times 10^{-4}$~ct~s$^{-1}$, from
which the absorption corrected flux and luminosity have been calculated
using the W3PIMMS software from LEDAS. For $N_{H}= 2.2 \times
10^{21}$~cm$^{-2}$, $kT=0.36$~keV (Stevens \& Strickland, 1998) and our
assumed distance of $22$~Mpc, the respective values are $(11.1\pm 2.5)
\times 10^{-15}$~erg~s$^{-1}$~cm$^{-2}$ and $(21.7\pm 4.8) \times
10^{38}$~erg~s$^{-1}$.

\section{Discussion}

\subsection{Morphology of Mrk~33}

Recent H$\alpha$ imaging of Mrk~33 (M\'{e}ndez et al. 2000) has shown
the presence of at least three star-forming knots in the centre of the
galaxy.  These regions are distributed in a line from NW to SE with the
largest knot being in the centre and the two most northern peaks being
closest together.  These two northern star-forming regions are
unresolved in our optical images and have been labelled N in
Fig.~\ref{Blue} whilst the southern one has been labelled S.  The region
occupied by the three star-forming knots has a size $\sim 7.1{''}\times
1.2{''}$ ($380\times 180$~pc at our assumed distance of 22~Mpc)
equivalent in area to a circular region of radius $\sim 2.1{''}$ ($\sim
225$~pc).  These sizes correspond well to the inner region with $r <
4{''}$ that shows the excess light and to the region occupied by the
peaks of optical emission shown in Fig.~\ref{gslice}.  In addition to
this, the $R^{1/4}$ fit allows the determination of the half-light
radius and the effective surface brightness at this radius.  For the
blue data these values are $4.35{''}$ and 20.35~mag~arcsec$^{-2}$
respectively while the red data gives values of $5.71{''}$ and
20.04~mag~arcsec$^{-2}$.  As typical compact dwarfs can be reasonably
well fitted by an $R^{1/4}$ law while their diffuse counterparts fit an
exponential law best it seems that Mrk~33 is morphologically best
described as a compact dwarf elliptical galaxy.

The changes in both Mrk~33's ellipticity and position angle with radius
are in reasonable agreement with the results of Loose \& Thuan (1986)
and Sage et al. (1992).  Differences between our results and those of
Loose \& Thuan (1986) could have arisen due to the different methods
used in ellipse fitting.  Our fitting had the ellipse centre fixed on
the optical centre of the most northern peak of the two shown on our
blue image (Fig.~\ref{Blue}) whilst their fitting had the ellipse centre
$x$ and $y$ positions as free parameters.  The increase in ellipticity
seen for radii $\leq 10{''}$ in our results has been influenced by the
linear arrangement of the star-forming knots in the galaxy centre
running from NW to SE which will tend to elongate the ellipses.The
isophotal twist observed means that our line-of-sight to Mrk~33 is not
along any of the galaxy's principle axes.  If it were, the isophotes
would be aligned like those of an axisymmetric system.  It is possible
that the twist is the result of the galaxy being triaxial but in no way
conclusive.  This could only be confirmed if the two true axial ratios
could be determined.

\subsection{Colour Gradient and Stellar Populations}

The $B-R$ colour profile shown in Fig.~\ref{ell} shows as expected that
the galaxy is very blue in the centre and becomes increasingly redder
with radius.  This reddening is indicative of a change in stellar
populations with increasing distance from the centre and is expected
from the population synthesis analysis of {\sl IUE} short-wavelength
spectra performed by Fanelli et al. (1988).  This analysis showed
distinct discontinuities in the stellar luminosity function of Mrk~33
with no significant levels of O7--B0V, B2--B9V or supergiants found.
The remaining stellar groups of O3--O6V, B1-B1.5V and A0-A7V were
detected in increasing numbers and this was interpreted as evidence for
Mrk~33 having experienced two episodes of intense star-formation prior
to the current one.  From the typical ages of the main sequence stars
detected, the current burst of star-formation must be around 5~Myr old,
the previous one occurred around 20~Myr ago and the earliest one around
500~Myr ago.  The fact that the number of stars detected in each group
is decreasing with time is indicative of the fact that the amplitude of
each successive starburst is also decreasing with time and may suggest
that there is less material from which to form stars left after each
intense star-forming episode.  At a radius $\sim 3.5{''}$, which would
just enclose the star-forming regions, $(B - R)\sim 0.83$, a colour that
is typical of an F5V star, and by the time a radius of $\sim 16{''}$ is
reached, this has increased to and levelled out at $B - R\sim 1.1$,
typical of a G0V star (Johnson 1966).  If these are the oldest stars
present then Mrk~33 only started forming stars $\sim 8\times 10^{9}$ yr
ago.

\subsection{Model of a Superbubble}

The extended X-ray emitting region in Mrk~33, appears to lie within the
expanding H$\alpha$ shell (diameter $\sim 2.8$~kpc) observed by Legrand
et al. (1997) in the centre of the galaxy, suggesting the presence of a
superbubble.  The model for such a starburst driven outflow (Castor et
al. 1975; Weaver et al. 1977) is outlined below.

Stars of type B2 and earlier have strong stellar winds associated with
them, which will deposit $\sim 10^{50}$~ergs of mechanical energy into
the ISM, during each star's lifetime. This is of the same order of
magnitude as the energy in a supernova shell.  As the winds from the OB
association of the starburst move out into the interstellar medium, they
will sweep up a thin, dense H$\alpha$ emitting shell.  Just inside the
shell will be a transition region that is dominated by thermal
conduction between the cold shell and the hot ($10^{7}$~K) shocked
stellar wind inside it.  The ionizing of hydrogen is likely to be
occurring in the shell itself and outside it there may be a layer of
H$_{2}$.  At first, the wind expands freely into the interstellar medium
(ISM) and shocks develop at its leading edge, a strong outward facing
shock moving at a speed close to the terminal velocity of the wind and a
weaker inward facing shock.  As more material is swept up into the
region between the two shocks, the momentum of the wind is insufficient
to maintain the high speeds and so the shell slows down.  This in turn
leads to the strength of the inner shock increasing because of the
increasing difference between the wind speed and the speed of the
shocked material.  This free expansion phase is very short and ends when
the mass of swept up material becomes greater than the mass deposited in
the region between the two shocks by the wind.  At this time the gas
between the shocks will be very hot with a temperature $\sim 10^{7}$~K
produced by collisions in the shocks converting the mechanical energy of
the wind to thermal energy.  This thermalisation is particularly
efficient because at these temperatures there is little cooling due to
radiative losses.

The superbubble then enters its adiabatic phase where it develops an
onion-like structure and can be thought of as consisting of five layers
(Strickland \& Stevens 1999).  Layer 1: - A freely expanding supersonic
wind bounded by the inward facing shock. Layer 2: - The shocked wind
material at $T\sim 10^{7}$~K.  Layer 3: - A contact discontinuity that
forms a boundary between the shocked wind material and shocked ISM
material.  Layer 4: - The hot swept-up shocked ISM material that is
bounded by the outward facing shock.  Layer 5: - The ambient ISM.
Adiabatic cooling occurs as work is being done increasing the bubble's
volume and when the point is reached where layer 4's temperature has
dropped to $\sim 10^{6}$~K, the emission of line radiation becomes
important and it's temperature rapidly falls to $\sim 10^{4}$~K.  The
two phases outlined above are very short, $\sim 10^{-5}\times $ the
lifetime of a starburst and so any observed superbubbles are much more
likely to be in their third phase which is known as the `snow-plough
phase'.

Layer 2 now occupies most of the bubbles volume and is at a uniform
pressure since it is very hot and its sound crossing time is very much
less than the age of the bubble.  Meanwhile, layer 4 which contains most
of the bubble's mass is now a thin, dense, cool shell with its
temperature maintained at $\sim 10^{4}$~K by the UV radiation from the
massive stars of the starburst.  The bubble continues to expand because
its pressure is greater than that of the surrounding ISM and its radius,
$R_{B}$, and expansion velocity $v_{B}$, are given by:

\begin{equation}
R_{B} = 32.77 \left (\frac {L_{36}}{\mu n_{0}} \right )^{1/5} 
t_{6}^{3/5}\;\;\;{\rm pc} 
\end{equation}

\begin{equation}
v_{B} = 19.28 \left(\frac {L_{36}}{\mu n_{0}}\right)^{1/5}
t_{6}^{-2/5}\;\;\;{\rm km\;s}^{-1}
\end{equation}

\noindent where $L_{36}$ is the mechanical luminosity injected by the
starburst in units of $10^{36}$~erg~s$^{-1}$ (the mechanical luminosity
injected by a typical B2 star), $t_{6}$ is the age of the starburst in
units of $10^{6}$ yr and $n_{0}$ is the ambient number density of the
ISM.

Thermal conduction at the contact discontinuity leads to the evaporation
of material from layer 4 to layer 2 (Weaver et al., 1977), where the
balancing of the conductive and mechanical energy fluxes lead to
expressions for the temperature, $T_{2}$ and the gas density, $n_{2}$,
at the centre of layer 2 given by:

\begin{equation}
T_{2} = 1.77 \times 10^{6}\;\; \left ( L_{36}^{8}\; n_{0}^{2}\; t_{6}^{-6}
\right )^{1/35}\;\;\;{\rm K}
\end{equation}

\begin{equation}
n_{2} = 0.02 \left ( L_{36}^{6}\; n_{0}^{19}\; t_{6}^{-22} \right ) ^{1/35}
\;\;\;{\rm cm}^{-3}
\end{equation}
 
Assuming the radiative cooling time is long compared to the expansion
time, spherical symmetry, a uniform density ambient ISM and a constant
rate of kinetic energy injection then:

\begin{equation} 
L_{X} = \int n(r)^{2}\Lambda_{X}(T,Z)dV\;\;\;{\rm erg~s}^{-1} 
\end{equation}
(Chu \& Mac Low 1990) where $\Lambda_{X}(T,Z)$ is the volume emissivity
of the gas over the temperature region being considered. Adopting
$\Lambda_{X} \sim 3\times 10^{-23}Z$~erg~cm$^{3}$~s$^{-1}$ for the
integral of the volume emissivity over the {\sl ROSAT} HRI energy band
(Suchkov et al., 1994; Raymond \& Smith, 1977), assuming $r = R_{B}$
given by equation (1) (since the cool shell is very thin compared with
the X-ray emitting region) and taking the radial variation of the
density inside the bubble, $n(r)$, to be:
\begin{equation} 
n(r) = 1.525 L_{mech}^{6/35} n_{0}^{19/35} t^{-22/35} \left(1 - \frac{r}{R} 
\right )^{-2/5}\;\;\;{\rm cm}^{-3}
\end{equation} 
(Mac Low \& McCray, 1988, following Weaver et al., 1977): where
$L_{mech}$ is the mechanical luminosity injected by the starburst and
$r/R$ is the fractional radius. Then, integrating over the total bubble
volume, $L_{X}$ is given by:
\begin{equation} 
L_{X} = 7 \times 10^{34} Z L_{36}^{33/35} n_{0}^{17/35} t_{6}^{19/35}
\;\;\;{\rm erg~s}^{-1}
\end{equation}

\subsection{Application of the model to Mrk~33}

To apply the ideas outlined above to Mrk~33, both an estimate of the
starburst age and mechanical energy injection rate are required. Legrand
et al.'s (1997) measurement of the size and outflow velocity of the
H$\alpha$ shell give values of $r\sim 1.3 - 1.4$~kpc (scaled to a
distance of 22~Mpc) and $v=200$~km~s$^{-1}$.  The earlier work of
Lequeux et al. (1995) studying Mrk~33's absorption lines of Ly$\alpha$,
OI, NI, SiII, and SiIII also found evidence for an outflow from the
galaxy, which was interpreted as a galactic wind with a velocity of
$\sim 200$~km~s$^{-1}$. Adopting this value for the expansion velocity
of the superbubble and assuming an average radius of $R_{B}=1.1$~kpc
(see Section 3.2.1) for the extended X-ray emitting source, a first
estimate of the dynamical age, $t$ can be made by taking the ratio of
equations (1) and (2):

\begin{equation}
t_{6} = 0.59\frac{R_{B}({\rm pc})}{v_{B}({\rm km~s}^{-1})}\;\;\;10^{6}{\rm yr} 
\Rightarrow t
= 3.2 \times 10^{6}{\rm yr} 
\end{equation}

\noindent This dynamical age would have to be a lower limit for the
starburst age and is most likely an underestimate of it since the
non-thermal nature of the radio spectrum implies a substantial number of
supernova remnants, suggesting that the starburst age should lie between
about 4~Myr and 40~Myr.  If, as is possibly indicated by the high
expansion velocity, the superbubble has already ruptured then this low
value for the dynamical age is a result of the hot gas and bubble shell
being accelerated to higher velocities during blow out. Once the bubble
has expanded to a radius of the order of the galaxy scale height, its
growth along the galaxy's minor axis will start to accelerate leading to
the rupture of the superbubble and blow out.  If the superbubble has a
radius of 1.1~kpc then it has already grown to several scale heights,
assuming the scale height is of the order of the half-light blue radius
($\sim 0.48$~kpc), and it is likely that acceleration is underway and
that blow out has just occurred.

Substituting the value for the dynamical age back into equation (1),
while assuming an average number density for the ambient ISM of
$n_{0}=0.3$~cm$^{-3}$ (following Marlowe et al., 1995, who quote this as
a typical value for this parameter in dwarf galaxies with recent or
on-going star-formation) and 90 per cent H to 10 per cent He, gives a
maximum value for the mechanical energy injection rate of:
$L_{36}=2.4\times 10^{5}$, or $L_{mech}=2.4\times
10^{41}$~erg~s$^{-1}$. The $L_{36}$ figure would suggest that the
maximum number of stars that could be present in the starburst would be
$\sim 2.4\times 10^{5}$, assuming they were all of type B2 and the
minimum mass of the starburst would be $\sim 2.35\times
10^{6}M_{\odot}$.

Comparing these values to the evolutionary synthesis models for an
instantaneous starburst (Leitherer \& Heckman 1995; Leitherer et
al. 1999) with a normal Salpeter IMF with $\alpha = 2.35$, an upper mass
cut-off of $100M_{\odot}$, a lower mass cut-off of $1M_{\odot}$ and
allowing for the fact that a metallicity of $0.3Z_{\odot}$ is being
assumed for Mrk~33, this energy injection rate is higher than would be
expected for a $10^{6}M_{\odot}$ starburst at $3.2$~Myr and suggests the
mass of the starburst is $\sim 2.9\times 10^{7}M_{\odot}$.  A better
quantity to consider instead of just the mechanical energy injection
rate, $L_{mech}$, is the ratio of $L_{mech}$ to $L_{bol}$, where
$L_{bol}$ is the bolometric luminosity of the starburst assumed to be
$\sim L_{FIR}$ (it is better to use this ratio because it is a mass
independent quantity).  Assuming $L_{FIR}\sim 1.4\times
10^{43}$~erg~s$^{-1}$, then $L_{mech}\sim 1.6\times
10^{40}$~erg~s$^{-1}$ at an age of $3.2$~Myr and the mass of the
starburst region $\sim 1.9\times 10^{6}M_{\odot}$.

Another estimate of the mass of the star-forming region of Mrk~33 has
been made by comparing the spectral energy distribution (SED) of the
galaxy to those predicted from SED modelling including single and binary
systems (Cervino \& Mas-Hesse 1999).  This gave an average value of
$\sim 9.5\times 10^{6}M_{\odot}$, assuming an =age of 5~Myr for the
starburst.  The justifications behind this choice are that Mrk~33 is a
Wolf-Rayet galaxy (Kunth \& Joubert 1985) suggesting an age of $3 -
7$~Myr.  Coupled with this, the Ly$\alpha$ profile of Mrk~33 (Lequeux et
al. 1995) is typical of that expected for a 5~Myr old starburst from the
evolution models of Tenorio-Tagle et al. (1999).  A final mass estimate
can be obtained from the number of O3--O6~V stars in the starburst. From
Fanelli et al. (1988), the extinction-corrected observed flux requires
$\sim 7150$ O3--O6~V stars (when scaled for different assumed
distances). Assuming a Salpeter IMF with an upper mass limit of
$120M_{\odot}$ to accommodate the O3 stars, this gives a mass of $\sim
9.3\times 10^{6}M_{\odot}$.  From the above considerations, a mass of
$10^{7}M_{\odot}$ is going to be assumed for the mass of the
star-forming region of Mrk~33 and all other quantities will be scaled
accordingly.

An upper limit on the age of the starburst is suggested by the $B-R$
colour of Mrk~33 in its central region.  The value of 0.56 shown in
Fig.~\ref{ell} would correspond to an age of $7.5$~Myr from the
evolutionary synthesis models but the actual value of $B - R$ in the
central region is likely to be bluer than this due to the presence of
large numbers of O and B stars and the fact that the surface brightness
has not been determined using elliptical isophotes that just include the
star-forming regions.

The flux in the H$\beta $ line found in Mrk~33's spectrum is
predominantly produced by the massive ionizing stars found in the
star-forming regions.  The equivalent width of this line, which
decreases with time as the number of hot stars decreases, is therefore a
measure of the relative number of these stars in the galaxy and hence
also an age indicator.  For Mrk~33, typical values of 21\AA\ (Kennicutt
1992), 23\AA\ (Conti 1991) and 30\AA\ (Mas-Hesse \& Kunth 1999) have
been quoted and these would correspond to ages of $\sim 6$, $5.9$ and
$5$~Myr respectively, from evolutionary synthesis models. It is possible
that these could be underestimates though as no allowance has been made
for the presence of binary stars in the galaxy, which would tend to
increase the age at which a particular equivalent width of H$\beta $
occurs at least for starbursts older than 4~Myr (Van Bever \& Vanbeveren
1998).  Also, the total number of photons below the Lyman limit can be
used as an age estimator. For Mrk~33, this quantity has a value of
$2.5\times 10^{52}$~photons~s$^{-1}$ (Conti 1991) which would correspond
to an age of $\sim 4.5$~Myr.

Table~1 gives the estimates of $L_{mech}$ calculated using equation (1)
and also those determined from the $L_{mech}$ to $L_{bol}$ ratio for the
six age estimates discussed above.  The values obtained from the
dynamical age differ by more than an order of magnitude and so both for
this reason and the reasons given earlier, these results will not be
included in the subsequent analysis.

\begin{table}
\caption{Estimates of the mechanical energy injection rate for
Mrk~33 for the six age estimates detailed in the text.  Column 2 gives the
value calculated by using equation (1) and a value for the radius of the
superbubble of 1.1~kpc.  Column 3 uses the $L_{mech}/L_{bol}$ ratios from the
starburst evolutionary synthesis models (Leitherer \& Heckman 1995;
Leitherer et al. 1999), assuming $L_{bol}$ for Mrk~33 to be $\sim 1.4\times
10^{43}$~erg~s$^{-1}$.}
\begin{tabular}{ccc} \hline
Age & Calculated $L_{mech}$ & $L_{mech}$ from $L_{mech}/L_{bol}$ \\ 
(Myr) & (erg~s$^{-1}$)    &  (erg~s$^{-1}$) \\ \hline
3.2 & $2.4\times 10^{41}$ & $1.6\times 10^{40}$ \\
4.5 & $2.2\times 10^{41}$ & $1.1\times 10^{41}$ \\
5.0 & $1.6\times 10^{41}$ & $1.1\times 10^{41}$ \\ 
5.9 & $9.6\times 10^{40}$ & $1.2\times 10^{41}$ \\
6.0 & $9.1\times 10^{40}$ & $1.1\times 10^{41}$ \\
7.5 & $4.8\times 10^{40}$ & $1.5\times 10^{41}$ \\ \hline
\end{tabular}
\end{table}

The average age and mechanical energy injection rate from the remaining
estimates are $5.8\pm 1.0$~Myr and $(1.2\pm 0.2)\times
10^{41}$~erg~s$^{-1}$ respectively. Using these values in equations (1)
and (2) give $R_{B} = 1.4\pm 0.2$~kpc and $v_{B} = 139\pm
18$~km~s$^{-1}$ respectively, which are in reasonably good agreement
with the observations. The fact that the expected expansion velocity is
less than that observed supports the idea that acceleration may be
occurring.  Compared with the evolutionary synthesis models, the figures
would suggest a mass for the starburst of $\sim 3.8\times 10^{6}M_{\odot
}$.  Another and probably more reliable estimate of the mass of the
starburst region can be obtained from comparing the absolute blue
magnitude of the starburst region with the values predicted from the
evolutionary synthesis models (Leitherer et al. 1999).  Assuming the
starburst is confined to the central region within a radius of $10^{''}$
then aperture photometry gives an apparent blue magnitude, $m_{B} =
13.8$ and an absolute blue magnitude, $M_{B} = -17.9$ for this region.
This value of $M_{B}$ is $\sim 2.1$ magnitudes brighter than predicted
for a $10^{6}M_{\odot}$ starburst at an age of $5.8$~Myr and implies
that the mass of the starburst is $\sim 6.9\times 10^{6}M_{\odot}$.
From this result, the assumption of a mass of $10^{7}M_{\odot}$ for the
starburst is not unreasonable.

Utilising equations (3) and (4) above along with the average age and
mechanical energy injection rates just determined, the temperature and
number density at the centre of layer 2 of the superbubble would be:
$T_{2}=(1.77\pm 0.19)\times 10^{7}$~K and $n_{2}=(0.026\pm
0.004)$~cm$^{-3}$ respectively assuming $n_{0}\sim 0.3$~cm$^{-3}$ as
stated earlier.  This temperature is somewhat higher than that obtained
from the spectral fitting of Stevens \& Strickland (1998) where a value
of $kT = 0.36$~keV was quoted.  This is to be expected however as the
conductive evaporation and mixing of denser gas from layer 4 with the
more tenuous gas of layer 2 just behind the contact discontinuity will
cause $T$ to drop rapidly here since $nT$ is constant.  The temperature
$T_{2}$ on the other hand corresponds to the temperature of the gas at
the centre of layer 2 just after it has been shock heated and assumes
100 per cent efficiency in the thermalizing of the mechanical energy
represents a maximum value for the X-ray temperature.

The model would predict that the X-ray luminosity, given by equation (7)
and assuming a metallicity of $0.3 Z_{\odot }$ for Mrk~33 should be:
$L_{X}=(1.8\pm 0.4)\times 10^{39}$~erg~s$^{-1}$.  This value is
comparable to that measured for the X-ray luminosity of the extended
central source.

\subsection{Other Sources of X-ray Emission}

The discussion in Section~4.4 shows that the extended X-ray source in
the centre of Mrk33, when treated as a whole, can be modelled as a
superbubble.  However, as this source has a blobby structure to its
emission, there are other possibilities and as in all galaxies, the
X-ray emission seen in Mrk~33 will come form a combination of sources
which will contribute in varying degrees to the overall observed flux.

Massive O stars will typically contribute around
$10^{31}-10^{32}$~erg~s$^{-1}$each to the X-ray luminosity (Sciortino et
al. 1990) giving around $10^{35}-10^{36}$~erg~s$^{-1}$ in total, while
lower mass stars with $L_{X}\sim 10^{-3}L_{Bol}$ (Stocke et al. 1991)
will contribute a similar amount. In total this is $< 1 \%$ of the
emission from the extended central source. X-ray binaries while giving
an adequate luminosity would appear as point sources and also would have
a harder spectrum than is observed in Mrk33 (Read, Ponman \& Strickland
1997). Supernova remnants have the right temperature spectrum (Read,
Ponman \& Strickland 1997) and the steep radio spectrum of Mrk33 is
indicative of the presence of such sources (Klein et al 1984). However,
the size of the extended source is $\sim 10 - 100$ times the size of
typical supernova remnants and several hundred would be required in a
very small volume to give the necessary luminosity. The short lived
nature of Type II supernovae and the need for them to occur in very
dense mediums (Schlegel 1995) or close to superbubble shell walls (Chu
\& Mac Low 1990) to produce the required luminosity means that they too
are unlikely to be responsible for most of the observed X-ray emission.

From the above considerations it seems most likely that the main
contribution to the X-ray emission from the extended central source is
from shock-heated gas within one or more superbubbles.  Improved spatial
resolution observations with say the Chandra HRC would be useful in
order to decide whether or not there are any point-like sources in the
region.

\subsection{The Long Term Effects on Mrk~33}

The fate of a dwarf galaxy experiencing a starburst depends on what
happens to the accelerated material from its ISM and that ejected from
its stars.  In essence, there are three possible outcomes to the
superbubble scenario.

Firstly, the ambient ISM could be accelerated to a speed beyond the
escape velocity of the galaxy and expelled along with the mass ejected
in the form of stellar winds and supernova ejecta.  Modelling carried
out by De Young \& Heckman (1994) shows that the most likely situation
to result in total loss of the ISM is a thick, low-density disk being
subjected to a high energy injection rate.  For a $10^{9} M_{\odot }$
galaxy with a semi-major axis of 3~kpc and an impulsive energy injection
of $10^{55}$~erg, they find that the ISM will only be lost if it has a
density of less than 0.1~cm$^{-3}$ and a semi-minor axis grater than
1~kpc.  These figures are not dissimilar to those of Mrk~33 but its ISM
is probably denser and although the total kinetic energy injected during
the starburst is $\sim 10^{55}$~erg, this energy has not been injected
impulsively hence, not all of the ISM is likely to be lost in the
current starburst.  This is supported by estimates of the escape
velocity for material at the current radius of the shell of the
superbubble.  Modelling the potential of the galaxy as a simple
spherically symmetric truncated isothermal potential (Binney \& Tremaine
1987), as has been done by several authors when modelling dwarf galaxies
(e.g. Marlowe et al. 1995; Martin 1999), the escape velocity for
material at a distance $r$ from the centre is given by:

\begin{equation}
v_{esc}(r) = 2^{1/2}v_{rot}[1 + ln(r_{t}/r)]^{1/2}\;\;\;{\rm km~s}^{-1}
\end{equation}
\noindent where $v_{rot}$ is the maximum rotation velocity of the galaxy in
km~s$^{-1}$ and $r_{t}$ is the radius at which the potential of the galaxy is
truncated.

In the case of Mrk~33, data for the rotation curve is unavailable and
the extent of its HI halo is also unknown, but some estimates can be
made based on data for other dwarf galaxies. A minimum distance for
$r_{t}$ can be set at the distance where the optical luminosity of the
galaxy appears to drop to that of the sky, i.e. $\sim 50{''}$ or
5.5~kpc. The work of Viallefond \& Thuan (1983), reviewing
interferometric maps of BCDGs suggests that their HI diameters could be
up to 5 times larger than their optical diameters. Setting this as an
upper limit gives $r_{t} = 27.5$~kpc. Substituting these values into
equation (9) for material at a distance of 1.1~kpc form the galaxy
centre gives:

\begin{equation}
2.28v_{rot} \leq v_{esc}(r) \leq 2.90v_{rot}\;\;\;{\rm km~s}^{-1}
\end{equation}

\noindent For the swept-up shell of ISM, expanding at 200~km~s$^{-1}$,
to escape, the maximum rotation velocity must be less than
70~km~s$^{-1}$. This seems a little lower than most of the rotation
speeds for the sample of galaxies discussed by Martin (1999).  The
acquisition of an HI rotation curve for this galaxy would be
particularly useful for determining both the HI extent and escape
velocity.

The second and more likely outcome for Mrk~33 is that a blow-out will
occur where the bubble ruptures due to the onset of Rayleigh-Taylor
instabilities allowing the hot metal-enriched gas to escape from the
fragmented shell.  Any further hot gas injected by the starburst can
then vent directly out of the galaxy, provided its sound speed is
greater than the escape velocity.  In the absence of radiative cooling,
gas hotter than $T_{esc} = 1.5\times 10^{5} (v_{100})^{2}$~K will escape
the galaxy's potential (Martin 1999).  Where $T_{esc}$ is the
temperature required for the gas to exceed the galaxy's escape velocity
and $v_{100}$ is the escape velocity in units of 100~km~s$^{-1}$.
Applying this to the gas in layer 2 of the superbubble, at a temperature
of $\sim 1.8\times 10^{7}$~K, shows that this gas could escape provided
$v_{esc} < 1100$~km~s$^{-1}$, which will undoubtedly be the case.  This
then raises the question of when the superbubble in Mrk~33 is likely to
become unstable.  The perturbations which give rise to the fractures
need time to grow and this growth occurs on a timescale given by Heckman
et al. (1995) as:

\begin{equation}
t_{R-T} \sim \sqrt {\frac {\Delta r_{bubble}}{2\pi g}} \sim  \sqrt {\frac
{\Delta r_{bubble}t_{exp}}{2\pi v_{B}}}
\end{equation}

\noindent where in this case, $g$ is the net acceleration of the shell
(the bubble expansion velocity divided by the expansion time), and
$\Delta r_{bubble}$ is the thickness of the shell.  Taking a typical
shell thickness of 0.1~kpc, this gives a time of 6.7~Myr for Mrk~33 and
suggests that bubble rupture is likely relatively soon and may already
have occurred as this value is within our error estimates on the age of
the starburst. This scenario is also supported by the work of Mac Low \&
McCray (1988). They define a criteria for blowout that is based on the
parameter $\Lambda$, the dimensionless rate of kinetic energy injection
given by:

\begin{equation}
\Lambda = 10^{4} L_{mech, 41} H_{kpc}^{-2} P_{4}^{-3/2} n_{0}^{1/2}
\end{equation}

\noindent where $L_{mech, 41}$ is the mechanical energy luminosity in
units of $10^{41}$~erg~s$^{-1}$, $H_{kpc}$ is the galaxy scale height in
kpc and $P_{4}$ is the initial pressure of the ISM in units of $P/k =
10^{4}$~K~cm$^{-3}$.  The condition specified for blowout is that
$\Lambda > 100$. Approximating $H_{kpc}$ to the half-light blue radius
and $P_{4} \sim 1$ then for Mrk~33, $\Lambda \sim 3\times 10^{4}$, thus,
blowout would appear to be extremely likely.

The third option is that the superbubble could decelerate and stall
resulting in radiative cooling and dissipation.  If the onset of
Rayleigh-Taylor instabilities is as imminent as the above calculations
would suggest then there will not be enough time for this last scenario
to occur since the winds of massive stars and their supernova explosions
will still be occurring to maintain the excess pressure and continued
expansion of the superbubble.

As blow-out seems the most likely, it is worth considering how much mass
has been deposited in the superbubble and hence how much is likely to be
lost from the galaxy when the shell ruptures.  A first estimate of the
mass deposition rate, $\dot {M}$ for a $10^{7}M_{\odot}$ starburst, with
$Z = 0.3 Z_{\odot}$ and $t = 5.8$~Myr can be obtained from the starburst
evolutionary synthesis models (Leitherer \& Heckman 1995; Leitherer et
al. 1999), and gives $\dot {M} \sim 0.17M_{\odot}$~yr$^{-1}$.  Another
estimate can be obtained from the fact that for an adiabatic
superbubble, the X-ray temperature, $T_{X} \sim 5/11T_{0}$, where

\begin{equation}
T_{0} = \frac {2}{3} \left ( \frac {L_{mech}}{\dot {M}} \right ) \frac
{\mu m_{H}}{k}\;\;\; {\rm K}
\end{equation}

\noindent (Heckman et al. 1995) and is the temperature attained if all
the mechanical energy injected by the starburst remains entirely in
layer 2.  For $T_{X}\sim 1.77\times 10^{7}$~K, a value of $\dot {M} \sim
0.24 M_{\odot}$~yr$^{-1}$ is obtained, in reasonably good agreement with
the first estimate. Hence the average mass loss rate from the starburst
in the centre of Mrk~33 is $\dot {M} \sim 0.2M_{\odot}$~yr$^{-1}$.

Finally, the density of the gas in layer 2 can give an estimate of the
mass deposited there during the lifetime of the superbubble from both
the starburst and conductive evaporation of ISM material from layer
4. This figure will of course depend on the filling factor of layer 2,
but assuming no dense clouds are present giving a filling factor is 1,
an upper limit for the deposited mass can be obtained. A second
assumption is that the radius of the X-ray emitting region is equivalent
to the radius of layer 2, $R_{2}$, and that any volume occupied by
freely expanding wind is very small compared to the volume of layer
2. Hence, with all symbols having their usual meanings and the subscript
2 referring to values within layer 2, 

\begin{equation} M_{2} = \rho_{2}
V_{2} = n_{2}\mu m_{H} \frac {4}{3} \pi R_{2}^{3} = 2.2\times
10^{6}M_{\odot} 
\end{equation} 

\noindent This mass has been deposited in a time of $5.8\times
10^{6}$~yr and so for a constant injection rate would imply $\dot {M}
\sim 0.38M_{\odot}$~yr$^{-1}$.  If the mass loss rate from the starburst
is assumed to be $\sim 0.2M_{\odot}$~yr$^{-1}$ then this would imply
that the rate at which mass is deposited in layer 2 by conductive
evaporation of ISM material from layer 4 is $\sim
0.18M_{\odot}$~yr$^{-1}$.  If the superbubble was to rupture at around
$6.7\times 10^{6}$~yr then assuming the mass is deposited at a constant
rate from both sources, the superbubble would contain a total mass of
$\sim 2.5\times 10^{6}M_{\odot}$, which would be expelled from the
galaxy in the form of a galactic wind.  Assuming this mass loss rate
continues for at least $10^{7}$~yr then around $4\times 10^{6}M_{\odot}$
of material could be lost from Mrk~33.  While this is of the same order
of magnitude as the mass of the starburst region, it is only $\sim 0.5 $
per cent of the total mass of the galaxy.  It appears likely then that
Mrk~33 can survive this starburst episode, just as it has survived its
previous two, with most of its ISM intact for use in subsequent
star-formation.

\subsection{Similar Stories in Other Dwarf Galaxies}

Mrk~33, like NGC~5253 is a dwarf starburst where the extended soft X-ray
emission seems to be confined well within the galaxy's optical extent
and within the observed H$\alpha $ emission.  These two galaxies have
probably not yet experienced blow-out of their current superbubbles.  In
contrast, NGC~1569 shows spurs of X-ray emission (Heckman et al. 1995)
which are coincident with its H$\alpha $ emission, a sign which is
indicative of a blow-out having occurred. The age estimate for NGC~1569
is $\sim 10^{7}$~yr and so it would appear to be an older and more
evolved starburst than either Mrk~33 or NGC~5253. NGC~1569 shows
evidence for several large kpc-scale structures containing diffuse X-ray
emission while Mrk~33 and NGC~5253 have extended regions of X-ray
emission that have a blobby structure.  All three of these galaxies show
evidence for having several distinct star-forming regions present in
their starbursts and each region will be responsible for a superbubble.
The combination of several superbubbles in this way will give rise to
the extended, complex, shell-like morphologies observed in these
galaxies.  The fact that the superbubbles in NGC~1569 are more easily
distinguished from each other than the blobs in the other two galaxies
is to be expected if it is older and has therefore had more time to
develop its superbubbles.

\section{Summary and Conclusions}

In summary, we have presented a detailed X-ray/optical analysis of the
important dwarf elliptical starburst galaxy Mrk~33. We find an extended
and complex X-ray morphology associated with the starburst region, and
we interpret this emission as being due to a superbubble or outflow
being inflated by the starburst. Consideration of the dynamics of the
situation show that Mrk~33 is a fairly young starburst galaxy with an
age of $t = (5.8\pm 1.0)\times 10^{6}$~yrs and that the rate of
injection of kinetic energy from the starburst region responsible for
inflating the superbubble is $L_{mech} = (1.2\pm 0.2)\times
10^{41}$~erg~s$^{-1}$.  The mass of the starburst region is estimated to
be $\sim 7\times 10^{6}M_{\odot}$. In the superbubble, the temperature
of the X-ray emitting gas at the centre of layer 2 is $T = (1.77\pm
0.19)\times 10^{7}$~K while its density is $n = 0.026\pm
0.004$~cm$^{-3}$.  Assuming a filling factor of 1 for this gas, we
estimate that the mass deposited in the superbubble from both the
starburst region and conductive evaporation of the ISM at the contact
discontinuity is $M \sim 2.2\times 10^{6}M_{\odot }$ and for the age
given above, this gives a mass injection rate of $\dot {M} =
0.38M_{\odot }$~yr$^{-1}$ of which, $\sim 50$ per cent comes from the
stars of the starburst in the form of stellar winds and supernova ejecta
and the remainder is the result of conductive evaporation of the
swept-up ISM.  It seems most likely that Mrk~33's superbubble will
become Rayleigh-Taylor unstable and rupture (in fact the high expansion
speed seen in this galaxy may indeed indicate that this has already
occurred). The resulting venting of the hot material from within it in
the form of a galactic wind will remove $\sim 0.5$ per cent of the total
mass of the galaxy and inject both energy and metal-rich material into
the intergalactic medium while allowing Mrk~33 to retain enough of its
ISM to continue star-formation in the future.

\section*{Acknowledgements}

LKS and IRS acknowledge funding from a PPARC studentship and Advanced
Fellowship respectively. DKS is supported by NASA through {\it Chandra}
Postdoctoral Fellowship Award Number PF0-10012, issued by the {\it
Chandra} X-ray Observatory Center, which is operated by the Smithsonian
Astrophysical Observatory for and on behalf of NASA under contract
NAS8-39073. The data analysis was performed on the {\sl Starlink} node
at Birmingham University.  This research has made use of data obtained
from the Leicester Database and Archive Service at the Department of
Physics and Astronomy, Leicester University, UK.  This research has also
made use of the SIMBAD database, operated at CDS, Strasbourg,
France. The JKT data analysis made use of the IRAF package, which is
distributed by NOAO, which is operated by AURA, Inc., under cooperative
agreement with the NSF.

\appendix
\section{Point Sources.}

Table~A1 lists all 23 sources detected with a significance greater than
$4\sigma $ above the background in the {\sl ROSAT} HRI field of view
during the 47.7~ks observation of Mrk~33.  The table is set out as
follows: Column 1 gives the source numbers that identify the sources on
Fig.~\ref{ImXC}; Column 2 gives the R~XJ name of each source according
to the {\sl ROSAT} naming convention (Zimmermann et al. 1997); Columns 3
and 4 contain the right ascension (J2000) and declination (J2000) of the
sources respectively where these positions are only accurate to within
$10{''}$; Column 5 lists the significance of each detection in terms of
the number of $\sigma $ above background; Column 6 gives the count rate
for each source in units of $10^{-4}$~ct~s$^{-1}$; and Column 7 gives
the hardness ratio ($HR =$ counts in channels $(6-8)/(3-5)$) for each
source.
 
\begin{table*}
\caption{The R~XJ numbers, positions, detection significance, count
rates and hardness ratios of the 23 sources detected in the {\sl ROSAT} HRI 
field of view during the 47.7~ks observation of Mrk~33. }
\begin{tabular}{ccccccc} \hline
Source & R~XJ number & RA (2000) & Dec. (2000) & Sig & Count
Rate & HR \\ 
 &  & (h m s) & ($^\circ$ $'$ $''$) & ($\sigma $ above background) & ($10^{-4}$~ct~s$^{-1}$)
&  \\ \hline
X-1 & 103102.0+542352 & 10 31 02.0 & $+54$ 23 52 & 7.6 & $13.8\pm 1.7$ & $0.42\pm 0.12$  \\
X-2 & 103130.2+542101 & 10 31 30.2 & $+54$ 21 01 & 5.9 & $5.7\pm 1.1$ & $0.36\pm 0.13$ \\
X-3 & 103136.1+541843 & 10 31 36.1 & $+54$ 18 43 & 6.8 & $6.9\pm 1.2$ & $0.59\pm 0.16$ \\
X-4 & 103151.8+542518 & 10 31 51.8 & $+54$ 25 18 & 6.6 & $6.1\pm 1.1$ & $0.43\pm 0.13$ \\
X-5 & 103152.0+541728 & 10 31 52.0 & $+54$ 17 28 & 4.5 & $3.9\pm 0.9$ & $0.54\pm 0.18$ \\
X-6 & 103153.1+541724 & 10 31 53.1 & $+54$ 17 24 & 4.6 & $3.9\pm 0.9$ & $0.42\pm 0.14$ \\
X-7 & 103155.4+541702 & 10 31 55.4 & $+54$ 17 02 & 7.1 & $6.7\pm 1.2$ & $0.50\pm 0.16$ \\
X-8 & 103156.6+543036 & 10 31 56.6 & $+54$ 30 36 & 4.4 & $2.8\pm 0.8$ & $0.65\pm 0.21$ \\
X-9 & 103205.5+543726 & 10 32 05.5 & $+54$ 37 26 & 4.2 & $7.5\pm 1.3$ & $0.38\pm 0.13$ \\
X-10 & 103209.1+543013 & 10 32 09.1 & $+54$ 30 13 & 9.0 & $9.2\pm 1.4$ & $0.44\pm 0.12$ \\
X-11 & 103219.4+541234 & 10 32 19.4 & $+54$ 12 34 & 5.2 & $6.2\pm 1.1$ & $0.69\pm 0.20$ \\
X-12 & 103219.5+543806 & 10 32 19.5 & $+54$ 38 06 & 5.0 & $7.9\pm 1.3$ & $0.73\pm 0.23$ \\
X-13 & 103229.2+542124 & 10 32 29.2 & $+54$ 21 24 & 4.8 & $3.9\pm 0.9$ & $0.78\pm 0.25$ \\
X-14 (Mrk~33) & 103232.0+542402 & 10 32 32.0 & $+54$ 24 02 & 5.9 & $5.1\pm 1.0$ & $0.11\pm
0.05$ \\
X-15 & 103233.6+542621 & 10 32 33.6 & $+54$ 26 21 & 4.8 & $3.3\pm 0.8$ & $0.32\pm 0.12$ \\
X-16 & 103244.0+543606 & 10 32 44.0 & $+54$ 36 06 & 8.8 & $14.3\pm 1.7$ & $0.23\pm 0.07$ \\
X-17 & 103249.0+543409 & 10 32 49.0 & $+54$ 34 09 & 14.0 & $21.6\pm 2.1$ & $0.12\pm 0.05$ \\
X-18 & 103249.3+541244 & 10 32 49.3 & $+54$ 12 44 & 5.0 & $5.4\pm 1.1$ & $0.69\pm 0.21$ \\ 
X-19 & 103251.5+541837 & 10 32 51.5 & $+54$ 18 37 & 4.2 & $2.6\pm 0.7$ & $0.46\pm 0.15$ \\
X-20 & 103322.8+542523 & 10 33 22.8 & $+54$ 25 23 & 12.2 & $13.8\pm 1.7$ & $0.21\pm 0.06$ \\
X-21 & 103341.9+541755 & 10 33 41.9 & $+54$ 17 55 & 4.1 & $3.6\pm 0.9$ & $0.53\pm 0.19$ \\
X-22 & 103349.7+541535 & 10 33 49.7 & $+54$ 15 35 & 6.6 & $10.8\pm 1.5$ & $0.74\pm 0.20$ \\
X-23 & 103402.6+543019 & 10 34 02.6 & $+54$ 30 19 & 6.0 & $11.8\pm 1.6$ & $0.35\pm 0.11$ \\
\hline
\end{tabular}
\end{table*}

\end{document}